\documentclass[aps,prb,twocolumn,showpacs,amsmath,amssymb,superscriptaddress,floatfix]{revtex4-2}
\usepackage[pdftex]{graphicx,hyperref}
\usepackage{hyperref}
\hypersetup{colorlinks = true, urlcolor = blue, linkcolor = blue, citecolor = blue}

\begin{document}

\title{Nonsymmorphic Symmetry Protected Dirac, M\"obius, and Hourglass Fermions in Topological Materials}

\author{Rui-Xing Zhang}
\affiliation{Department of Physics and Astronomy, University of Tennessee, Knoxville, Tennessee 37996, USA}
\affiliation{Department of Materials Science and Engineering, University of Tennessee, Knoxville, Tennessee 37996, USA}
\affiliation{Institute for Advanced Materials and Manufacturing, University of Tennessee, Knoxville, Tennessee 37996, USA}
\author{Chao-Xing Liu}
\email{cxl56@psu.edu}
\affiliation{Department of Physics, the Pennsylvania State University, University Park, PA 16802}

\begin{abstract}
A lattice symmetry, if being nonsymmorphic, is defined by combining a point group symmetry with a fractional lattice translation that cannot be removed by changing the lattice origin. Nonsymmorphic symmetry has a substantial influence on both the connectivity and topological properties of electronic band structures in solid-state quantum materials. In this article, we review how nonsymmorphic crystalline symmetries can drive and further protect the emergence of exotic fermionic quasiparticles, including Dirac, M\"obius and hourglass fermions, that manifest as the defining energy band signatures for a plethora of gapless or gapped topological phases of matter. We first provide a classification of energy band crossings in crystalline solids, with an emphasis on symmetry-enforced band crossings that feature a nonsymmorphic-symmetry origin.
In particular, we will discuss four distinct classes of nonsymmorphic-symmetry-protected topological states as well as their signature fermionic modes: (1) a $Z_2$ topological nonsymmorphic crystalline insulator
with two-dimensional surface Dirac fermions; (2) a Dirac semimetal with three dimensional bulk-state Dirac nodes pinned at certain high
symmetry momenta; (3) a topological M\"obius insulator with massless surface modes that resemble the topological structure of a M\"obius twist (dubbed as M\"obius fermions); (4) a time-reversal-invariant version of M\"obius insulators with hourglass-shaped massless surface states (dubbed as hourglass fermions). The emergence of the above exotic topological matter perfectly demonstrates the crucial roles of both symmetry and topology in understanding the colorful world of quantum materials.
\end{abstract}

\maketitle

\section{Introduction}

Symmetry describes the invariance of a physical system under certain operations. Over the past century, symmetry has been playing a more and more fundamental role in our understanding of physical laws in nature. For example, Einstein's invention of both special and general relativity theories is based on the deep insight that space and time are symmetric. The recognition of the elegance of local gauge symmetry in Maxwell's equation of electromagnetism motivated Yang and Mills to generalize the gauge theory to the non-Abelian Lie groups\cite{yang1954conservation}, and this generalization eventually led to the unification of electromagnetic, weak and strong forces, forming the basis of the Standard Model of particle physics. In condensed matter physics, the symmetry principle is the key to classifying various phases of matter, as well as the phase transitions among them. Take magnetism as an example. Despite the isotropic microscopic interaction between magnetic moments, magnetic phase transitions can occur between paramagnetic and ferromagnetic states that possess different symmetries: paramagnetic (disordered) state has a full rotational symmetry, the same as the microscopic interaction, while ferromagnetic (ordered) state breaks the rotation symmetry. This observation reveals one of the most important concepts in condensed matter physics, the {\it spontaneous symmetry breaking}. As the temperature is reduced to a critical value, an order of magnetization (described by a mathematical quantity termed {\it order parameter}) is spontaneously generated to break rotation symmetry, leading to the ferromagnetic state. An effective field theory, known as the Landau-Ginzburg theory \cite{landau1980statistical}, can be mathematically formulated based on the order parameter to describe this spontaneous symmetry breaking mechanism. Remarkably, such a field theory can capture the universal properties of phase transitions between different states of matter in the sense that it only relies on certain general properties of a physical system, such as the dimensionality of the system and the symmetry of order parameters. The 20th century has witnessed the great triumph of the application of the symmetry principle in the development of almost all branches of theoretical physics. In the article "Symmetry and Physics" written by Prof. C.N. Yang in 1996 \cite{yang1996symmetry}, he asked: ``Will the next century witness new facets of the concept of symmetry in our deepening understanding of the physical world?" And for this question, he gave an affirmative answer,``I would answer, yes, very probably. "

Indeed, the development of condensed matter theory in the first two decades of 21st century has echoed Prof. Yang's optimistic anticipation. This time, symmetry cooperates with another mathematical concept, topology, and together leads to the discovery of new types of quantum states of matter, known as topological states, in condensed matter systems \cite{qi2011topological,hasan2010colloquium}. Two topologically distinct states cannot be adiabatically connected to each other without a topological phase transition that is usually signaled by an energy gap closing. Different from normal phase transitions, these two states may share the same symmetry property. Thus, topological states and topological phase transitions cannot be classified by the usual symmetry principle in the Landau-Ginzburg paradigm, and instead, topological invariants, i.e., a discrete-valued mathematical quantity, are responsible for characterizing different topological states. Due to its discreteness nature, a (topological) phase transition has to occur due to the variation of this topological invariant. The quantum Hall effect, which occurs in a two-dimensional (2D) electron gas system under a strong magnetic field and was discovered by Von Klitzing in 1980 \cite{klitzing1980new}, gives the first example of topological states. The quantum Hall states are characterized by the Chern number, a topological invariant that is an integral of the momentum-space Berry curvature for the whole Brillouin zone of a solid, a quantity that can be constructed through the electron wave-functions. Remarkably, the Chern number in a quantum Hall state is directly related to the Hall conductivity\cite{thouless1982quantized}, which can be measured in standard magneto-transport experiments. Another salient feature of topological states is the certain type of gapless boundary modes, of which the existence originates from the bulk topological property instead of local boundary conditions. For example, chiral edge states, which only propagate in a certain direction, exist at the edge of a quantum Hall system, and carry dissipationless edge currents. These exotic boundary modes of topological states hold great potential in ushering in new electronic devices with low-energy consumption and extraordinary functionalities.

The quantum Hall state had remained to be the only known topological state for almost two decades and its existence does not rely on any symmetry. Around 2005, researchers started to realize that symmetries can greatly enrich the topological structures of quantum solids. The discovery of topological insulators gives us the first example of time-reversal-symmetry-protected topological states \cite{qi2011topological,hasan2010colloquium}. 
Unlike the quantum Hall state that requires external magnetic fields to break time-reversal symmetry (TRS), topological insulators exist in non-magnetic materials with strong spin-orbit coupling, and its topological nature is characterized by a $Z_2$ topological invariant\cite{kane2005z}, which means that there are only two types of insulators, a trivial insulator with $Z_2$ number 0 and a non-trivial insulator with $Z_2$ number 1. It turns out that TRS is essential for defining this $Z_2$ topological invariant appropriately. Topological insulators also possess gapless boundary modes, but different from the chiral edge states that break TR symmetry in the quantum Hall system, the gapless boundary modes of topological insulators respect the Kramers theorem \cite{kramers1930theorie}
for time-reversal invariant systems. 
The Kramers theorem states that an eigenstate of a non-magnetic solid at the momentum ${\bf k}$ should have the same energy as its time-reversed eigenstate at the momentum $-{\bf k}$. These two states, normally called a "Kramers pair", possess opposite spins, and thus are orthogonal to each other. The Kramers theorem guarantees two salient features of the boundary modes of topological insulators. First, a double degeneracy has to occur at time-reversal invariant momenta, e.g. ${\bf k}=0$, and this double degeneracy is essential for protecting the gapless nature of boundary modes of topological insulators. If TRS is broken, this double degeneracy will be removed and the boundary modes of topological insulators will be gapped out. Second, the opposite spin for a Kramers pair requires the helical nature of the boundary modes of topological insulators, which means that the spin of the boundary modes is locked to the momentum and forms a texture in the momentum space. Thus, the boundary modes of topological insulators are dubbed "helical states".

\begin{figure*}[t]
    \includegraphics[width=0.75\textwidth]{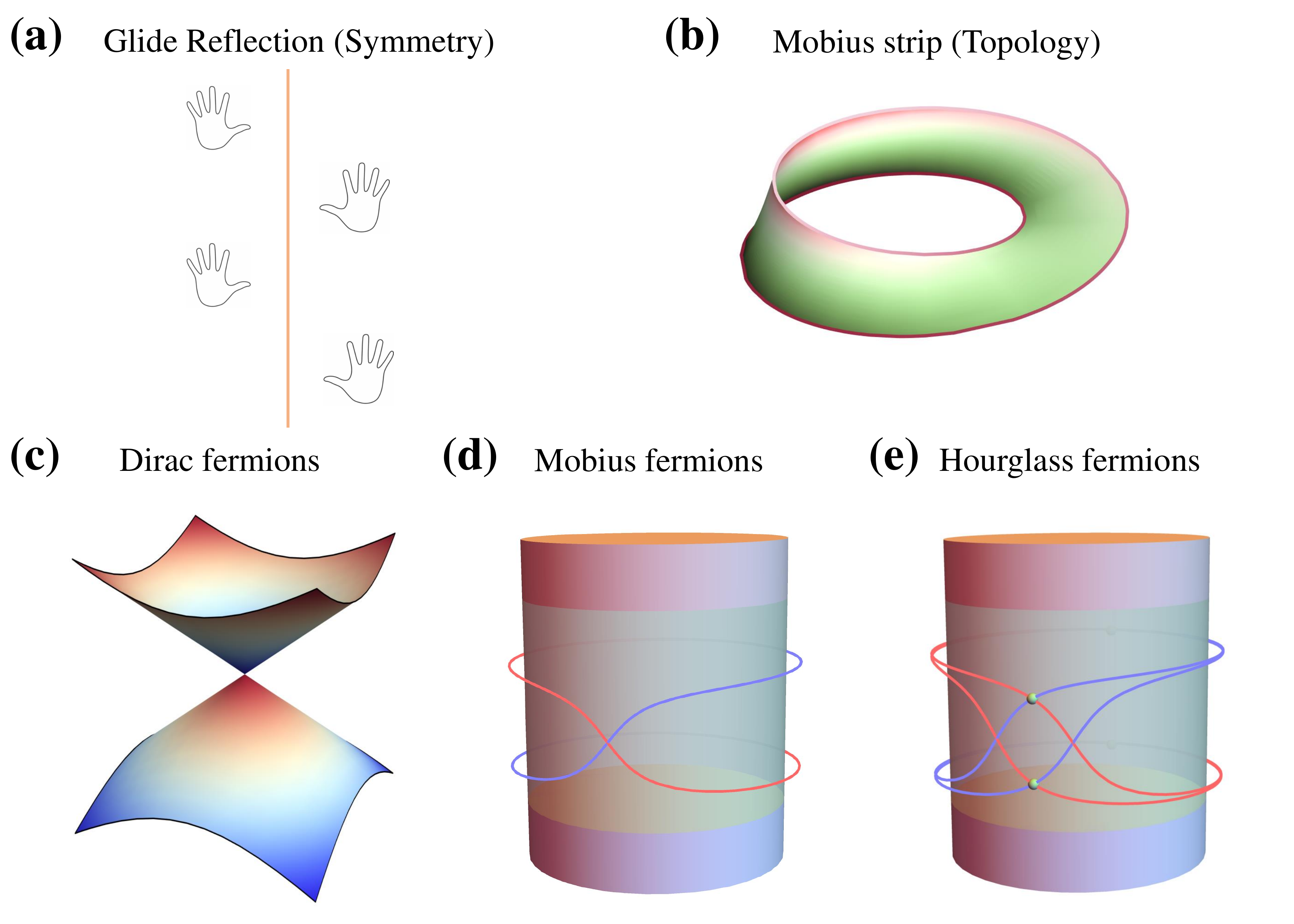}
	\caption{(a) Schematics for glide reflection symmetry. (b) M\"obius strip as a topological object. (c), (d) and (e) schematically show
		the forms of Dirac, M\"obius and hourglass fermions in the momentum space. The cylinders in (d) and (e) depict the periodic Brillouin zone. }
	\label{Fig1}
\end{figure*}

The most important lesson from the discovery of topological insulators is the central role of TRS. Thus, it is natural to ask if other types of symmetries can protect new topological states, and soon afterwards, researchers started to explore this question. These studies were extremely fruitful and led to an explosion of the whole field with the discoveries of a large group of new topological states. The first generalization to other internal symmetries, including particle-hole symmetry and chiral symmetry, led to a ten-fold way classification of topological states \cite{schnyder2008classification}. Particularly, the existence of particle-hole symmetry in the BdG formalism of superconductivity implies the possibility of topological superconductors \cite{kitaev2001unpaired,read2000paired}, in which the corresponding boundary modes, e.g. Majorana zero modes, may carry non-Abelian statistics and thus can serve as the basis for constructing topological quantum bits for quantum computation\cite{nayak2008non}. It was later realized that the crystalline symmetry in solids can also protect non-trivial boundary modes, and the richness of crystalline symmetry led to the rapid development of the field of topological crystalline insulators \cite{fu2011topological}. This development motivated people to re-examine the connection between the Bloch band representation of electronic band structure and the local atomic orbital basis, and eventually led to a systematical theoretical framework, including the symmetry indicator theory \cite{po2017symmetry} and topological quantum chemistry \cite{bradlyn2017topological}, to understand topological properties of
electronic band structure. It was also shown that the nodes, including nodal points, lines and planes, in semimetals are deeply connected to topological properties of electronic bands \cite{armitage2018weyl}, giving rise to the classification of topological semimetals, such as Weyl and Dirac semimetals, nodal line, ring and chain semimetals. Beyond non-interacting systems, the idea of topological phases protected by symmetry was also extended to interacting systems\cite{chen2012symmetry}, leading to even richer classifications. The realization of these topological states was not limited to electronic systems, but instead, some other systems, such as cold atoms, photonics, phononics, magnonics and mechanical systems, are actually more feasible for designing the model Hamiltonians for the realization of these new topological phases. The predictions and discoveries of new quantum phases and materials with their deep connection to both symmetry and topology are one of the most exciting developments in theoretical condensed matter physics in the first two decades of the 21st century.

In this review article, we will focus on a class of topological states that are related to a special type of crystalline symmetry, called nonsymmorphic symmetry, in this exciting field that connects topological states and symmetry. In a solid, crystal symmetry normally consists of two types of operations, namely translation and point group symmetry (e.g. reflection, rotation, etc). A nonsymmorphic symmetry refers to a symmetry operation combining a point group symmetry with a fractional lattice translation that cannot be removed by changing the origin.  
Typical nonsymmorphic symmetry operations include glide reflection [see Fig. \ref{Fig1} (a)] and screw axis
rotation. Nonsymmorphic symmetry operations strongly affect the form of electronic band structures in solid materials, namely they tend to "glue" different energy
bands together, forming band crossings. When these band crossings occur around Fermi energy, they can give rise to different exotic forms of fermions in the low-energy effective sector in the materials [see Fig. \ref{Fig1} (c) - (e)], which are deeply connected to the topological property of electronic band structure.
Below we will illustrate how nonsymmorphic symmetry enables new topological states and what types of topological states can be resulted in. To see that, we will first illustrate the relation between band crossings and nonsymmorphic symmetry in the next section.

\section{Band Crossing and nonsymmorphic symmetry}
In quantum mechanics, if two linearly independent eigenstates of a Hamiltonian share the same energy, we say these two states are energetically degenerate \cite{landau2013quantum}. Degeneracy is closely related to symmetry. For an eigen-state $|\alpha\rangle$ of a Hamiltonian, the state $|\alpha'\rangle=\hat{S}|\alpha\rangle$ created by a symmetry operation ${\cal S}$ is also an eigen-state of the Hamiltonian with the same energy.
If $|\alpha\rangle$ and $|\alpha'\rangle$ are linearly independent, they form a symmetry ${\cal S}$-enforced degenerate pair of states for the Hamiltonian, and such a symmetry-protected degeneracy is usually classified as {\it essential degeneracy}. In addition, there is another type of degeneracy, called {\it accidental degeneracy}, which occurs coincidentally without any protection by symmetry \cite{von1993verhalten}.

\begin{figure*}[t]
	\includegraphics[width=0.95\textwidth]{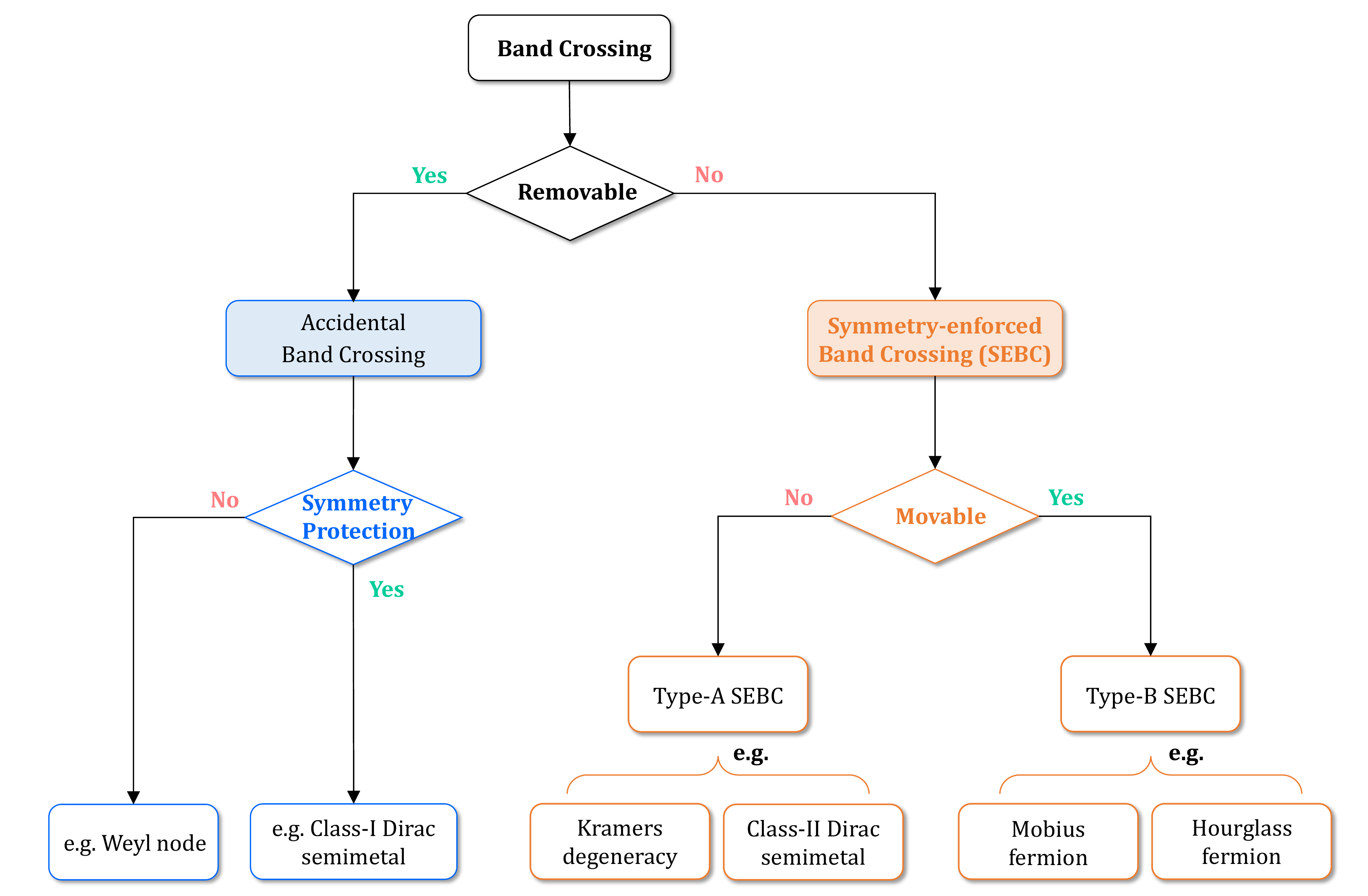}
	\caption{Classification of band crossings. Depending on the removability of band crossings, we classify band crossing as "accidental" and "symmetry-enforced". For symmetry-enforced band crossings, we further classify into type-A and type-B, depending on if the crossing is movable
		in the momentum space or not. For accidental crossings, we also classify them into two classes, depending on the role of symmetry. Examples of topological states that are related to each class of band crossings are given. }
	\label{Fig2}
\end{figure*}

Both types of degeneracies are ubiquitous in solid-state systems. In a solid, the energy states of electrons vary continuously with the crystal momentum and form energy bands in the Brillouin zone (BZ), the Wigner-Seitz primitive cell of the reciprocal lattice, due to the periodic potential provided by the lattice. Treating the crystal momentum as a parameter, different energy bands are generally split from each other, but they may become degenerate at certain momenta and these degenerate points are called ``band crossings". Similar to degeneracies, the band crossings in solids can be related to symmetry, namely the crystal symmetry that in mathematics is described by the space group and anti-unitary time-reversal symmetry, or they can be accidental. Below, we will classify two types of band crossings, namely the ``symmetry-enforced band crossing" (SEBC) and ``accidental band crossing" (ABC),
depending on if these band crossings are removable or not when a certain crystalline symmetry group is imposed, as shown in the second row of Fig. \ref{Fig2}.

SEBCs are guaranteed by crystal symmetries or time-reversal symmetry and cannot be removed by any symmetry-preserving perturbation.
We can further classify SEBCs into A-type and B-type, depending on the ``movability" of these band crossings in the momentum space
(as shown in Fig. \ref{Fig2}).
The type-A SEBC is related to higher-dimensional irreducible representation (IR) of a symmetry group.
In the BZ of a crystal, all the eigen-states of the Hamiltonian at certain crystal momentum can be classified by the IRs of the wave-vector group (or little group) at that momentum. If the IR of wave-vector group has the dimension more than 1, the eigen-states that correspond to these IRs must
be degenerate with its degeneracy given by the dimension of the IR. For example, as will be discussed in Sec. \ref{Sec_Diracsurface}, the surface Dirac point of a spinless topological nonsymmorphic crystalline insulator is essentially a 2-fold degeneracy that arises from the two-dimensional IR of the wavevector group in the surface BZ.
Since the higher-dimensional IR normally occurs at isolated high-symmetric momenta in the BZ, the resulting band crossings are pinned at these momenta.
As we will see below, higher-dimensional IR at high-symmetry momenta can appear in a variety of nonsymmorphic symmetry groups.
Anti-unitary symmetry operation, e.g. time-reversal symmetry, can also give rise to higher-dimensional IR, with the example of Kramers degeneracy at time-reversal invariant momenta in a non-magnetic crystal.
Type-B SEBC is not from high-dimensional IR, and instead it occurs between two different one-dimensional (1D) IRs at certain high symmetry lines
or planes. Unlike the pinning of type-A SEBC at isolated momenta, the type-B SEBC can move in the high-symmetry momentum lines or planes of the BZ although their existence is also guaranteed by crystal symmetry groups. Type-B SEBCs typically appear in nonsymmorphic symmetry groups and are closely related to the unique properties of nonsymmorphic symmetry operations, as discussed below.

We refer accidental band crossings (ABCs) to a class of band crossings that result from the Wigner-Von Neumann theorem of codimension theory\cite{von1993verhalten}. ABCs do not rely on the protection of any crystal symmetries and are generally movable in the momentum space. However, unlike SEBCs, ABCs must {\it come in pairs} in the whole BZ as a result of energy band connectivity. Consequently, ABCs are only {\it locally stable} in the sense that while a single ABC is stable by itself, it can always annihilate with its pair partner when they meet in k-space. Notably, the stability of ABCs can always be further enhanced by certain crystal symmetries or TRS. A well-known example of ABCs is 2-fold-degenerate {\it Weyl nodes} that can be found in three dimensional (3D) Weyl semimetals candidates such as TaAs. Here, the codimension for two energy levels to be degenerate is 3, so a Weyl node is locally stable in 3D momentum space, but a pair of Weyl nodes with opposite chiralities can annihilate with each other.
Unlike its 3D counterpart, a 2-fold ABC in 2D (i.e. a 2D Dirac point) is not even locally stable in general, while this local stability can be restored if certain symmetry (i.e. the combined symmetry of two-fold rotation and time-reversal) is present.
This type of ABC is quite similar to the type-B SEBC, as both can move in certain high-symmetry lines or planes in the momentum space,
but there is one essential difference: the role of crystal symmetry here is to change the codimension of nodal points to make such band crossing locally stable, while the type-B SEBC is globally stable and guaranteed by certain symmetry group.

\section{Nonsymmorphic symmetry protected topological states}
In this section, we will discuss the possible forms of exotic fermions in certain topological materials due to the band crossings
that originate from nonsymmorphic symmetry groups. These exotic fermions are closely related to topological properties
of the materials: they are either gapless boundary modes of topological insulating phases or bulk gapless modes in topological semimetals.
We will discuss three types of exotic fermions: Dirac, M\"obius and hourglass fermions, and focus on their close relationship to the type-A and type-B SEBCs
discussed above. Particularly, in Sec. \ref{Sec_Diracsurface} and \ref{Sec:DiracBulk}, we will discuss nonsymmorphic-symmetry-protected 2D and 3D Dirac nodes,
which result from the type-A SEBCs and manifest as the defining spectrum features for topological nonsymmorphic crystalline insulators (TNCIs) and class-II Dirac semimetals, respectively. In Sec. \ref{Sec:Mobius} and \ref{Sec:hourglass}, we will discuss M\"obius fermions and hourglass fermions, both of which manifest as type-B SEBCs with interesting band dispersion relations and live on the surface of glide-mirror-protected topological crystalline insulators.

\subsection{Type-A SEBC: Surface Dirac Fermions in Topological Nonsymmorphic Crystalline Insulators}\label{Sec_Diracsurface}

Type-A SEBC is related to the high-dimensional IR of the wave-vector group at certain momenta. Normally, the wave-vector group of a crystal momentum can
contain both one-dimensional and higher-dimensional IRs. What is special to nonsymmorphic symmetry group is that some high-symmetry momenta may
only host higher-dimensional IRs, due to the presence of two anti-commuting symmetry operators at these momenta. Let us consider two symmetry operators
$\hat{S}_1$ and $\hat{S}_2$ that anti-commute with each other, $\{\hat{S}_1,\hat{S}_2\}=0$. We choose the state $|\alpha,s_1\rangle$ as
the common eigen-state of the Hamiltonian and the symmetry operator $\hat{S}_1$, where $s_1$ is the eigen-value of $\hat{S}_1$, $\hat{S}_1|\alpha,s_1\rangle
=s_1|\alpha,s_1\rangle$. As a symmetry operator, the state $\hat{S}_2|\alpha,s_1\rangle$ is also an eigen-state of the Hamiltonian, but due to the anticommutation
relation, $\hat{S}_1\hat{S}_2|\alpha,s_1\rangle=-\hat{S}_2\hat{S}_1|\alpha,s_1\rangle=-s_1\hat{S}_2|\alpha,s_1\rangle$. which means that the state $\hat{S}_2|\alpha,s_1\rangle$ carries the eigen-value $-s_1$ under $\hat{S}_1$, opposite to that of $|\alpha,s_1\rangle$. Therefore, $|\alpha,s_1\rangle$ and
$\hat{S}_2|\alpha,s_1\rangle$ must be linearly independent and thus are two degenerate eigen-states. If this situation occurs at certain momenta in the BZ
of a crystal, the corresponding wave-vector group will only have higher-dimensional IRs and all the eigen-states must be at least two-fold degenerate.
This situation indeed occurs for the nonsymmorphic symmetry group. More interestingly, it is possible to have these type-A SEBCs as a topological boundary
for a TNCI, which was first pointed out by Liu {\it et al.} in Ref. \cite{liu2014topological} and will be discussed below.

\begin{figure*}[t]
	\centerline{\includegraphics[width=0.7\textwidth]{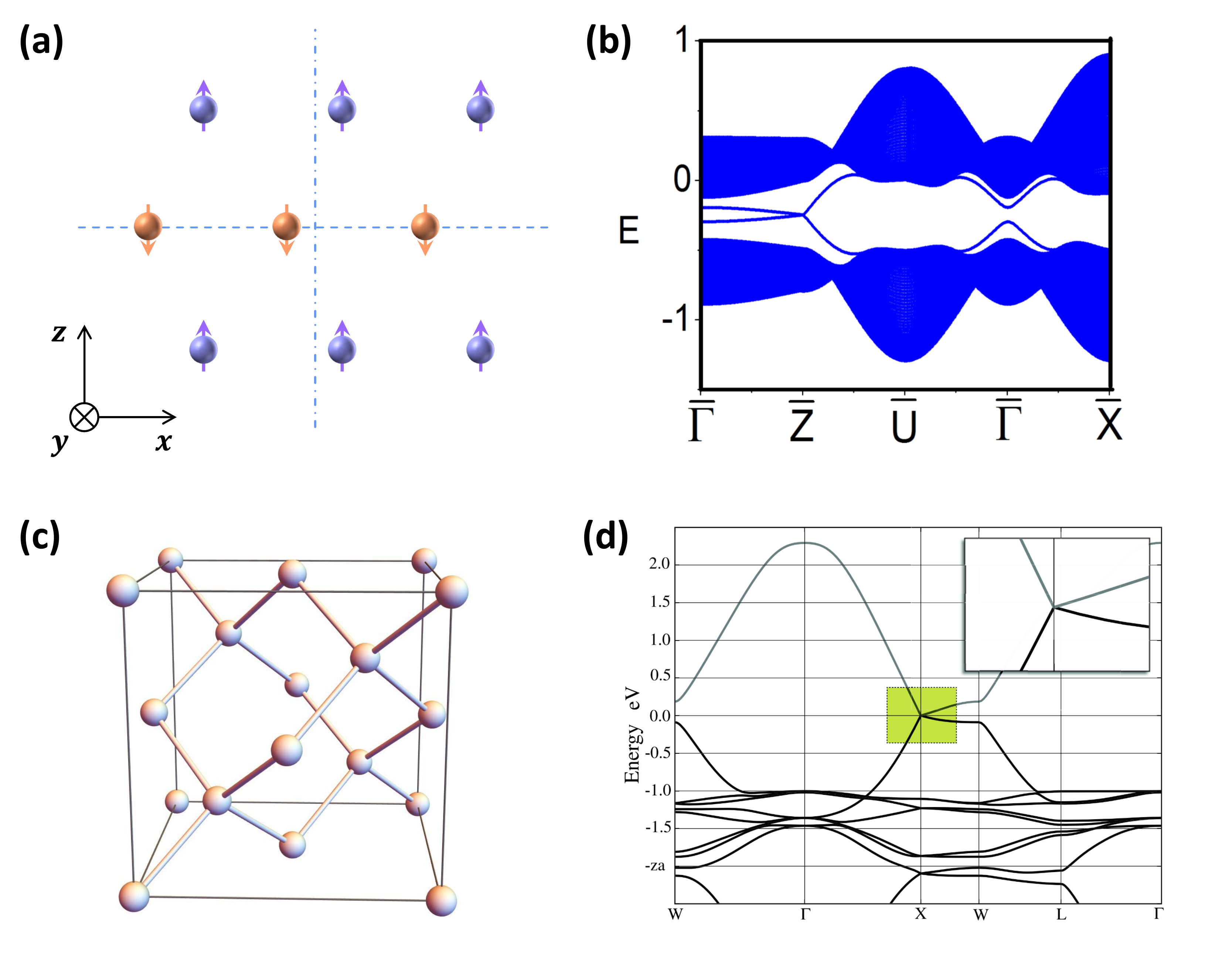}}
	\caption{(a) Schematics of the lattice for the model of topological nonsymmorphic crystalline insulators. (b) Surface state spectrum of topological
		nonsymmorphic crystalline insulators. From Ref.~\cite{liu2014topological}. (c) Diamond lattice. (d) Three-dimensional Dirac fermions in the electronic
		band struture of BiO$_2$. From Ref.~\cite{young2012dirac}.
	}
	\label{Fig3}
\end{figure*}

The TNCI phase can emerge in a 3D layered antiferromagnetic system,
with each layer being ferromagnetically ordered but adjacent layers carrying opposite out-of-plane magnetic moments.
As shown in Fig. \ref{Fig3}a, a principle layer of the lattice thus contains two oppositely polarized sub-layers, denoted as A and B. 
Notably, the A and B sublayers are further displaced along the x axis, namely, the position of the A atom is $\vec{r}_A = (-a_1,0,0)$ while that of the B atom is $\vec{r}_B = (a_1,0,c_2)$.
Here, the lattice vectors are denoted as $\vec{a}_1 = (a,0,0)$, $\vec{a}_2 = (0,b,0)$, and $\vec{a}_3 = (0,0,c)$.
This configuration has the space-group symmetry of $Pma2$-type, with two essential symmetry operations as the group generators,
the mirror symmetry along the z direction, denoted as $\hat{m}_z = \{\hat{m}_z|\vec{e}\} : (x,y,z) \rightarrow (x,y, -z)$
where $\vec{e} = (0,0,0)$, and the glide symmetry $\hat{g}_x = \{\hat{m}_x|\vec{\tau} \} : (x,y,z) \rightarrow (-x,y,z + \frac{c}{2})$
with $\vec{\tau} = \frac{\vec{a}_3}{2}= (0,0, c_2)$. These two symmetry operations do not commute with each other,
$\hat{m}_z\hat{g}_x=\{\hat{C}_{2y}|-\vec{\tau}\}\neq \hat{g}_x\hat{m}_z=\{\hat{C}_{2y}|\vec{\tau}\}$, where $\hat{C}_{2y}$
is a twofold rotation around the y axis. It is exactly this noncommutativity relation between $\hat{g}_x$ and $\hat{m}_z$ that give rise to
a 2D IR of the wave-vector group, which further leads to a SEBC as the topological surface Dirac fermion.

To demonstrate that, we consider three orbital degrees of freedom, i.e., one s-orbital (dubbed $\phi_s$) and two p-orbitals
(dubbed $\phi_{x}$ and $\phi_y$), at each lattice site. The linear combinations of $\phi_x$ and $\phi_y$ orbitals form
a set of doublet states that carry angular momentum $\pm1$ and thus can couple to magnetic moments through Zeeman-type coupling.
The basis wave-functions for this system can be written as $|\alpha \eta,\vec{k}\rangle=\frac{1}{\sqrt{N}}\sum_n e^{i\vec{k}\cdot\vec{R}_{n\eta}}\phi_{\alpha}(\vec{r}-\vec{R}_{n\eta})$, where $\vec{R}_{n\eta}=\vec{R}_n+\vec{r}_{\eta}$
with $\vec{R}_n$ is the lattice vector and $\eta=A, B$ for two sublattices.
The Hamiltonian of this system, denoted as $H_{\alpha\eta, \alpha'\eta'}(\vec{k})$,
can be written on these basis wave-functions and its detailed form can be found in Ref. \cite{liu2014topological},
which generally describes a six-band insulating state.
Acting the symmetry operators $\hat{m}_z$ and $\hat{g}_x$ on the basis wavefunctions lead to
$\hat{m}_z|\alpha\eta,\vec{k}\rangle=|\alpha\eta,\hat{m}_z\vec{k}\rangle$ and $\hat{g}_x|\alpha\eta,\vec{k}\rangle=\sum_\beta e^{-ik_zc/2}m_{x,\alpha\beta}|\beta\bar{\eta},\hat{m}_x\vec{k}\rangle$, where $\bar{\eta}$ is the interchange of A and B indices
and the 3$\times$3 matrix $m_x=\text{diag}(1,-1,1)$ in the basis of $\phi_s$, $\phi_x$ and $\phi_y$.
Then it is straightforward to verify that along two special high-symmetry lines $\vec{\kappa}_1=(0,k_y,\pi/c)$ and $\vec{\kappa}_1=(\pi/a,k_y,\pi/c)$ that remain invariant under both $\hat{m}_z$ and $\hat{g}_x$ operations, we have the following anticommutation relation
\begin{equation}
	\{\hat{g}_x,\hat{m}_z\}|\alpha\eta,\vec{\kappa}_{1,2}\rangle = 0.
	\label{eq:anticom_TNCI}
\end{equation}

As described above, the existence of two anti-commuting symmetry operators	immediately suggests all the eigen-states along the $\vec{\kappa}_{1,2}$ lines must
be doubly degenerate. Now let us denote $|\Psi^{I}_{\vec{\kappa}}\rangle$ as the common eigen-state of $H(\vec{\kappa}_{1,2})$ and $\hat{m}_z$.
Due to the anti-commutation relation Eq. (\ref{eq:anticom_TNCI}),
$|\Psi^{II}_{\vec{\kappa}_{1,2}}\rangle = e^{i\chi_{\kappa_{1,2}}}\hat{g}_x |\Psi^{I}_{\vec{\kappa}_{1,3}}\rangle$ is another linearly independent degenerate eigen-state, where $\chi_{\kappa_{1,2}}$ is a phase shift. With these two sets of degenerate states $|\Psi^{I}_{\vec{\kappa}}\rangle$ and $|\Psi^{II}_{\vec{\kappa}}\rangle$ as the "doublet pairs", one can define
one $Z_2$ topological invariant $\Delta=P_d(k_x=\pi/a)-P_d(k_x=0) mod 2$ in the $k_z=\pi/c$ plane (formed by the parallel $\kappa_{1,2}$ lines),
where $P_d(k_x)=P_I(k_x)-P_{II}(k_x)$ with charge polarization $P_{\alpha}(k_x)=\frac{1}{2\pi}\oint dk_y\langle\Psi^{\alpha}|i\partial_{k_y}|\Psi^\alpha\rangle$,
in analogous to the time-reversal-protected $Z_2$ topological invariant for topological insulators \cite{fu2007topological}. This $Z_2$ topological invariant
characterizes the topological classification of the $Pma2$ group: $\Delta=1$ for the TNCI phase and $\Delta=0$ for the trivial insulator phase.
We emphasize that the doubly degeneracy enforced by the anti-commutation relation between $\hat{g}_x$ and $\hat{m}_z$ allows this $Z_2$ topological invariant to be well defined.

The TNCI phase with $\Delta=1$ supports 2D gapless Dirac fermion on (010) surfaces for a finite-size system.	
By imposing an open boundary condition along $\hat{y}$ direction while remaining periodic ones for $\hat{x}$ and $\hat{z}$ directions, a slab geometry with open $(010)$ surfaces can be constructed, in which both $\hat{g}_x$ and $\hat{m}_z$ are carefully preserved. Since $k_y$ is no longer a good quantum number, only two high-symmetry momenta $\bar{Z}=(k_x=0,k_z=\pi/c)$ and $\bar{M}=(k_x=\pi/a,k_z=\pi/c)$ can support the desired SEBCs for the surface BZ. The surface bands generally split when moving away from $\bar{Z}$ and $\bar{M}$, forming a linear Dirac-like dispersion (i.e. a 2D Dirac cone on the surface).
This scenario is supported by a direct calculation of the surface electronic band structure for the above $\hat{y}$-directional slab based on the model
Hamiltonian constructed in Ref. \cite{liu2014topological}. As shown in Fig. \ref{Fig3}b, the system features a single surface Dirac node at $\bar{Z}$ of
the surface BZ as expected, which explicitly confirms the $\mathbb{Z}_2$-nontrivial nature of the phase ($\Delta=1$).
Notably, the simultaneous presence of in-gap Dirac surface states at both $\bar{Z}$ and $\bar{M}$ will make both surface states unstable,
which corresponds to a topologically trivial phase. Therefore, the surface state analysis of the TNCI model also admits $\mathbb{Z}_2$ topological classification,
where the topologically nontrivial phase with $\Delta=1$ features an odd number of Dirac surface states.

\subsection{Type-A SEBC: Bulk Dirac Fermions in Class-II Dirac Semimetals}\label{Sec:DiracBulk}
The discussion above only involves two-fold degeneracy protected by nonsymmorphic crystal symmetry in a magnetic model system that breaks TRS.
In non-magnetic nonsymmorphic crystals, the cooperation between nonsymmorphic crystal symmetry and TRS can give rise to even higher degeneracy,
which has important applications in the robust bulk band crossings in 3D semimetals, especially those with point nodes.
In this part, we focus on the mechanism of 4-fold-degenerate 3D Dirac node in solid-state crystals, which manifests as a composite, relativistically dispersing quasiparticle made by ``gluing" a pair of oppositely charged Weyl fermions. Systems with robust and symmetry-protected Dirac nodes are termed ``Dirac semimetals" (DSMs)\cite{armitage2018weyl}. Thus far, two inequivalent types of 3D DSMs have been identified in 3D band systems, which we dub ``class-I DSMs" and ``class-II DSMs", respectively. Despite sharing similar Dirac nodal physics, class-I and class-II DSMs are clearly distinguished from each other in several important aspects:
\begin{itemize}
	\item {\it ABC v.s. SEBC}: (i) When two doubly degenerate bands cross, a pair of class-I Dirac nodes are formed only if the above bands carry distinct 2D IRs of the wavevector group so that no anti-crossing will happen. (ii) In contrast, a class-II Dirac node manifests itself as a 4D IR of certain nonsymmorphic space groups.
	\item {\it Number of Dirac nodes}: (i) Class-I Dirac nodes can only be created and annihilated pairwise, and a class-I DSM must contain an even number of Dirac nodes. (ii) Class-II DSM can host an odd number of Dirac nodes, as will be shown later.
	\item {\it Surface state}: Class-I DSMs can host topological surface states, while class-II DSMs generally do not.
\end{itemize}
Therefore, class-I Dirac nodes belong to movable yet symmetry-protected ABCs, while class-II Dirac nodes are neither removable nor movable and thus belong to type-A SEBCs, as summarized in Fig. \ref{Fig2}. In the following, we will show how nonsymmorphic symmetries can enable class-II Dirac nodes.

Even before the first theoretical proposal of 3D DSMs by Young {\it et al.}~\cite{young2012dirac}, a prototype of class-II DSM has already been studied in an earlier paper by Fu {\it et al.} (dubbed ``Fu-Kane-Mele" model or FKM model for short) \cite{fu2007topological2}, even though this paper is more well-known for discovering a nearby 3D topological insulating phase. The FKM model is defined on a 3D diamond lattice [as shown in Fig.~\ref{Fig3} (c)] and contains a pair of spinful s-orbitals per lattice site. Notable, the energy spectrum of the FKM model necessarily consists of a single 4-fold energy band degeneracy at each of the three inequivalent $X^{(r)}$ points at the BZ boundary, where $X^{(r)}=\frac{2\pi}{a}\hat{r}$ and $r\in\{x,y,z\}$ and $a$ is the lattice constant.
Such 3D Dirac nodes are clearly unmovable due to the specialty of the little group at $X^{(r)}$, thus manifesting as type-A SEBCs.

To understand the SEBCs in the FKM model, we note that the diamond lattice is under the space group symmetry of {\it Fd$\bar{3}$m}. The lattice itself can be intuitively understood by stack two identical copies of face-centered-cubic (FCC) lattices and shift them by a partial translational vector ${\bf t}_d = \frac{a}{4}(1,1,1)$ along the diagonal direction, as shown in Fig. \ref{Fig3}c. 
Three crucial symmetries are responsible for inducing the four-dimensional (4D) IR at $X^{(r)}$: (i) time-reversal symmetry (TRS) $\Theta$; (ii) spatial inversion symmetry $\tilde{\cal I}=\{{\cal I}|{\bf t}_d\}$ with ${\cal I}{\bf r}\rightarrow -{\bf r}$; (iii) two-fold rotation symmetry $C_{2z}$ around [001] axis.
It should be noted that although both $\tilde{\cal I}$ and $C_{2z}$ are point group symmetries, the origins for these two operations do not coincide
with each other, and the space group {\it Fd$\bar{3}$m} is nonsymmorphic. Below we will show that
this shift of the origins for two different symmetry operations can also lead to the anti-commutation relation at certain momenta.

\begin{figure*}[t]
	\centerline{\includegraphics[width=0.5\textwidth]{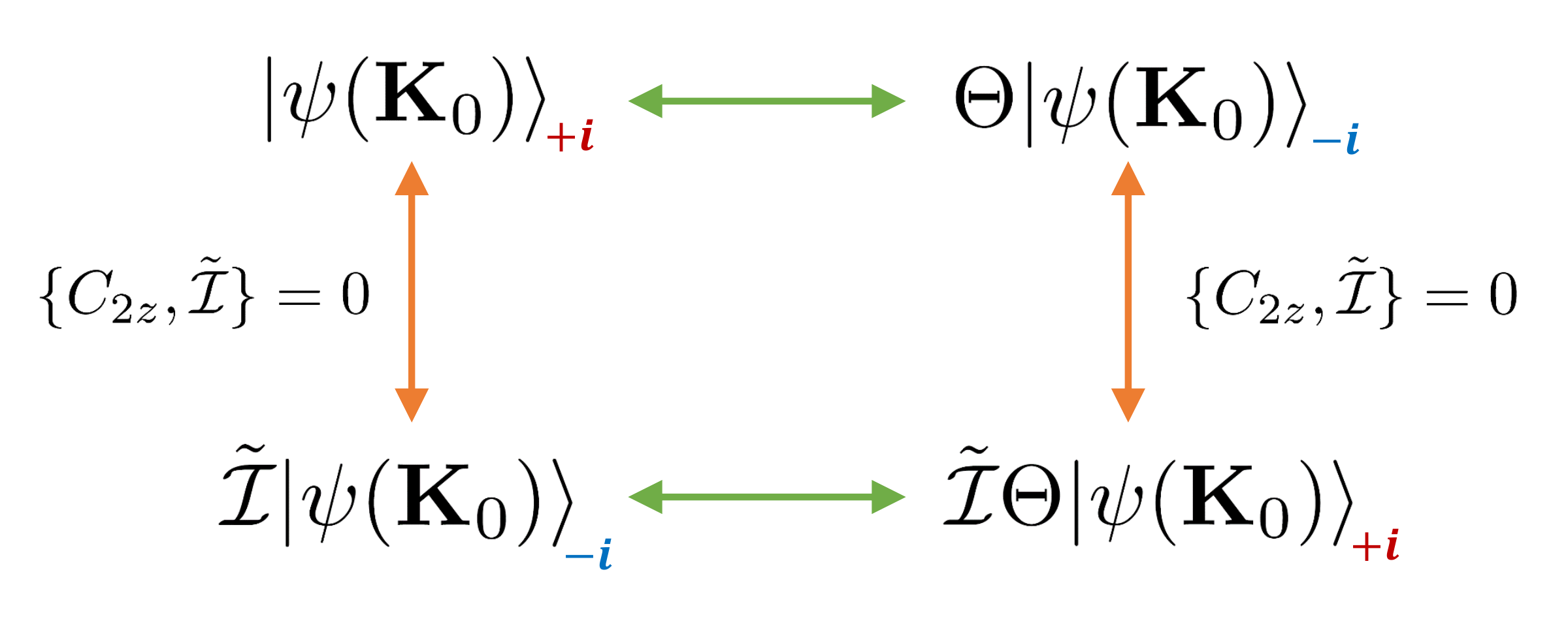}}
	\caption{Four degenerate eigen-states are related to each other by three symmetry operations $\tilde{\cal I}$, $C_{2z}$
		and $\Theta$. }
	\label{Fig:4D_IR}
\end{figure*}

While TRS commutes with all lattice symmetries, $\tilde{\cal I}$ and $C_{2z}$ do not generally commute since 
\begin{eqnarray}
	&&C_{2z}\tilde{\cal I}|\psi(k_x,k_y,k_z)\rangle = e^{i(k_x+k_y+k_z)\frac{a}{4}}C_{2z}{\cal I}|\psi(k_x,k_y,k_z)\rangle,  \nonumber \\
	&&\tilde{\cal I}C_{2z}|\psi(k_x,k_y,k_z)\rangle = e^{i(-k_x-k_y+k_z)\frac{a}{4}}C_{2z}{\cal I}|\psi(k_x,k_y,k_z)\rangle, \nonumber \\
	&&
\end{eqnarray}
where we have used the fact that $[C_{2z}, {\cal I}]=0$. Therefore, we arrive at
\begin{equation}
	C_{2z}\tilde{\cal I}|\psi(k_x,k_y,k_z)\rangle = e^{i\frac{a}{2}(k_x+k_y)}\tilde{\cal I}C_{2z}|\psi(k_x,k_y,k_z)\rangle.
\end{equation}
As a result, at high-symmetry momenta ${\bf K}_0 = X^{(x)}$ or ${\bf K}_0 = X^{(y)}$, it is easy to see that the following anticommutation relation is satisfied,
\begin{equation}
	\{C_{2z}, \tilde{\cal I}\}|\psi({\bf K}_0)\rangle = 0.
	\label{eq:anticom_DSM}
\end{equation}
This anti-commutation relation works for both spinful and spinless cases. We discuss here the spinful case and consider
a $C_{2z}$-eigenstate $|\psi({\bf K}_0)\rangle$ at ${\bf K}_0$ with
$C_{2z}|\psi({\bf K}_0)\rangle=+i|\psi({\bf K}_0)\rangle$.
Eq.~\ref{eq:anticom_DSM} immediately establishes $\tilde{\cal I}|\psi({\bf K}_0)\rangle$ as another eigen-state with the same energy yet an opposite $C_{2z}$ eigen-value.
Namely, $|\psi({\bf K}_0)\rangle$ and $\tilde{\cal I}|\psi({\bf K}_0)\rangle$ are doubly degenerate states. On the other hand, $\Theta|\psi({\bf K}_0)\rangle$, the Kramers partner of $|\psi({\bf K}_0)\rangle$, also features $C_{2z}=-i$. Despite carrying the same $C_{2z}$ index, $\tilde{\cal I}|\psi({\bf K}_0)\rangle$ and $\Theta|\psi({\bf K}_0)\rangle$ are necessarily distinct states due to $(\tilde{\cal I}\Theta)^2=-1$. Or equivalently, $\tilde{\cal I}\Theta|\psi({\bf K}_0)\rangle$, the Kramers partner of $\tilde{\cal I}|\psi({\bf K}_0)\rangle$ with $C_{2z}=+i$, should also be orthogonal to $|\psi({\bf K}_0)\rangle$. Therefore, the 4D IR at ${\bf K}_0$ is exactly formed by the following set of quantum states:
$\{|\psi({\bf K}_0)\rangle,\Theta|\psi({\bf K}_0)\rangle,\tilde{\cal I}\Theta|\psi({\bf K}_0)\rangle,\tilde{\cal I}|\psi({\bf K}_0)\rangle\}$, as summarized in Fig.~\ref{Fig:4D_IR}.

A similar argument can be formulated for the 4D IR at $X^{(z)}$ by considering the combination of $\Theta$, $C_{2x}$, and $\tilde{\cal I}$. Therefore, there are in total {\it three} 4-fold-degenerate Dirac nodes in the FKM model, and the Dirac nodes are further related to one another through a three-fold rotation symmetry along the [111] direction. It is worth noting that Ref.~\cite{yang2015topological} further pointed out that the class-II Dirac nodes in the diamond lattice is additionally protected by $\{\tilde{\cal I}, \tilde{C}_{4z}\}=0$, where $\tilde{C}_{4z}$ is a four-fold screw axis around [001] axis.

While the existence of such SEBC is a general feature for diamond-structure materials, searching for a promising candidate with Dirac nodes living near the Fermi level appears challenging. Ref. \cite{young2012dirac} reported $\beta$-cristobalite BiO$_2$ as a possible candidate with type-II DSM phase.
As shown in Fig.~ \ref{Fig3} (d), ab-initio-based band structure for BiO$_2$ looks very similar to the FKM model, which features large dispersing Dirac bands and class-II Dirac nodes at three inequivalent $X^{(r)}$s. However, Ref.~\cite{young2012dirac} also pointed out that $\beta$ BiO$_2$ calculated to be metastable structure. Therefore, it would be experimentally challenging to synthesize this desired structure, instead of Bi$_2$O$_4$ that is more favored energetically.

\subsection{Type-B SEBC: M\"obius Fermions}\label{Sec:Mobius}

We now proceed to discuss the more interesting situation of type-B SEBCs that are movable but irremovable. The possibly simplest realization of type-B SEBCs is known as the ``M\"obius fermions", which naturally occur in glide-mirror-symmetric magnetic systems. A M\"obius strip is the simplest non-orientable surface that can be created by giving one end of a paper strip a halftwist and further gluing the two ends together to form a closed loop, as shown in Fig. \ref{Fig1} (b). Being one-sided, a creature living a M\"obius strip can eventually run through both the ``front" and ``back" sides of the paper strip, without crossing any boundary of the strip.
In this part, we will demonstrate how a single glide mirror symmetry will necessarily enforce the energy bands to feature a similar M\"obius twist in its symmetry representation and how such M\"obius fermions can emerge as a new type of surface state for {\it higher-order M\"obius insulators} (HOMIs). The Physical origin of accompanied higher-order topology and material proposal of HOMIs will also be discussed.

To understand the origin of M\"obius bands, we consider a 3D TRS-breaking lattice system with a glide mirror symmetry $\hat{g}_x=\{\hat{m}_x|\hat{\tau}\}$
as in Sec. \ref{Sec_Diracsurface}, where the mirror reflection $\hat{m}_x:(x,y,z)\rightarrow (-x,y,z)$ and the half-lattice-translation $\hat{\tau}:(x,y,z)\rightarrow (x,y,z+c/2)$. For our purpose, we will focus on the surface M\"obius bands on the glide-invariant (010) surfaces for this 3D system, since the bulk counterparts can be understood analogously. Notably, the (010) surface BZ contains two glide-invariant lines (GILs) with $k_x=0/\pi$, which we denote as $\vec{\kappa}$. Along $\vec{\kappa}$, glide eigenvalue becomes a good quantum number and thus every surface state will be labeled by the glide index $g_x=\pm ie^{ik_z/2}$.

The M\"obius twist of the surface bands is a direct outcome of the {\it $4\pi$-periodicity of $g_x$} along $k_z$, even though the surface BZ is $2\pi$-periodic. To see this, we consider the evolution of an energy eigenstate $|\Phi_1(k_z)\rangle_{g_x}$ that carries a glide index $g_x=+ie^{ik_z/2}$. At $k_z=-\pi$, the energy and glide index of $|\Phi_1(-\pi)\rangle_+$ are given by $E_+$ and $g_x=+1$. After the eigen-state evolves continuously and smoothly across the surface BZ and arrives at $k_z=\pi$, however, the glide index for $|\Phi_1(\pi)\rangle_+$ now becomes $g_x=-1$. Therefore, the energies for $|\Phi_1(\pm\pi)\rangle_+$ do NOT need to coincide due to the flip of glide index, and we hence label the energy at $k_z=\pi$ as $E_-$. Due to the $2\pi$-periodicity of the BZ, there must also exist another  eigenstate at $k_z=-\pi$ with $g_x=-1$ and $E_-$, which we label as $|\Phi_2(-\pi)\rangle_-$. Notably $|\Phi_2(k_z)\rangle_{g_x}$ carries a glide index of $g_x=-ie^{k_z/2}$. Namely, going from $-\pi$ to $\pi$, $|\Phi_2(k_z)\rangle_{g_x}$ must evolve into an eigenstate at $k_z=\pi$ with both $E_+$ and $g_x=+1$ and further ``glue" with our starting state (i.e. $|\Phi_1\rangle_+$ at $k_z=-\pi$), in order to guarantee both the band connectivity and $2\pi$ periodicity. To summarize, the minimal set of glide-preserving energy bands necessarily consists of two bands along a GIL (i.e. $|\Phi_1\rangle$ and $|\Phi_2\rangle$), and they follow the following evolution path along $k_z$ to form a fully connected closed loop of band manifold:
\begin{widetext}
	\begin{equation}
		|\Phi_1(-\pi)\rangle_+^{(E_+)} \rightarrow |\Phi_1(\pi)\rangle_-^{(E_-)} \equiv |\Phi_2(-\pi)\rangle_-^{(E_-)} \rightarrow |\Phi_2(\pi)\rangle_+^{(E_+)} \equiv |\Phi_1(-\pi)\rangle_+^{(E_+)}.
	\end{equation}
\end{widetext}
Remarkably, the fact that $|\Phi_1\rangle$ and $|\Phi_2\rangle$ are {\it locally} distinct for a fixed $k_z$ but {\it globally} connected across the BZ exactly resembles the topological structure of a M\"obius strip. In analogy with the ``twist \& glue" trick for creating a paper M\"obius strip, the glide symmetry offers the ``twist" in the symmetry representation of the electron wavefunctions, while the periodicity of BZ serves as a natural ``glue" for the band manifold.

\begin{figure*}[t]
	\centerline{\includegraphics[width=0.99\textwidth]{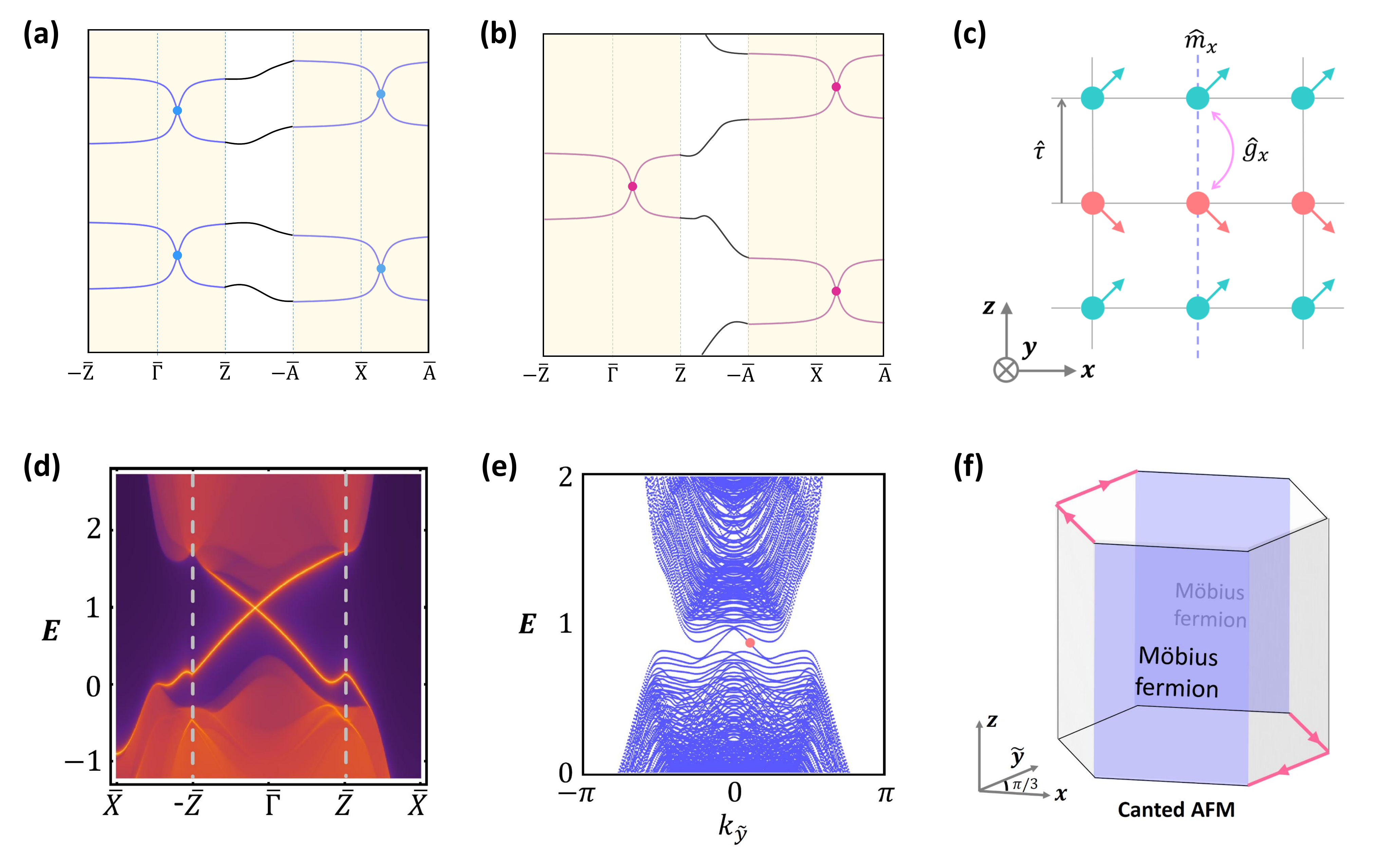}}
	\caption{(a) Schematic of a trivial, gapped M\"obius surface state. (b) Schematic of a topological, gapless M\"obius surface state that is equivalent to the one in MnBi$_{2n}$Te$_{3n+1}$. (c) Canted AFM phase in MBT$_n$ that preserves glide mirror symmetry $\hat{g}_x$. (d) 2D gapless M\"obius surface state of MBT$_n$ from the effective model calculation. (e) 1D gapless chiral hinge modes in MBT$_n$ along $k_{\tilde{y}}$ direction. (f) Connectivity of M\"obius surface states and chiral hinge modes in a hexagonal prism geometry of MBT$_n$ with canted AFM phase. Such an exotic mixed-dimensional topological boundary defines a higher-order topological M\"obius insulator phase. (c) - (f) are from Ref.~\cite{zhang2020mobius}}
	\label{Fig4}
\end{figure*}

Since $E_+\neq E_-$, $|\Phi_1\rangle$ and $|\Phi_2\rangle$ necessarily cross along GIL, while the location of band crossing is flexible or movable. Such a band crossing is protected by the distinct glide indices carried by the two crossed bands. The necessity, movability, and symmetry protection of the glide-mirror-enforced band crossing together establish it as an exemplar of type-B SEBCs.

Electronic insulators with surface M\"obius fermions as topological boundary modes are often known as ``M\"obius insulators". In particular, Ref. \cite{PhysRevB.91.155120} and \cite{PhysRevB.91.161105} have established a $\mathbb{Z}_2$ topological classification for M\"obius insulators. Namely, only systems with an odd number of surface M\"obius fermions are topologically nontrivial, while an even number of M\"obius  fermions can always be gapped and fail to contribute to topological surface conduction of electrons. The $\mathbb{Z}_2$ classification can be intuitively understood in the schematics in Fig. \ref{Fig4} (a) and (b). In particular, Fig.~ \ref{Fig4} (a) displays the scenario of two surface M\"obius fermions at a given Fermi energy, with each of them living along a GIL. Apparently, the connectivity between the two M\"obius fermions indicates the presence of a gapped surface and thus a trivial band topology, as the Fermi level always crosses an even number of Fermi surfaces. On the contrary, the Fermi level in Fig.~ \ref{Fig4} (b) always crosses an odd number of Fermi surfaces or M\"obius fermions. In particular, the zigzag connection between surface M\"obius bands at different GILs guarantees the presence of a robust gapless surface, which thus corresponds to a nontrivial band topology.

Since M\"obius fermions require the explicit breaking of TRS, the relevant material search in magnetic solids is challenging. Recently, Zhang {\it et al.} \cite{zhang2020mobius} proposed a feasible realization of M\"obius insulator in the newly discovered family of magnetic topological materials MnBi$_{2n}$Te$_{3n+1}$ (MBT$_n$), where $n=1,2,3,...$. Intuitively, MBT$_n$ can be understood as an alternate stacking of a MnTe layer and $n$ copies of Bi$_2$Te$_3$ layer, where the Bi$_2$Te$_3$ part contributes to a single topological band inversion at the center of bulk BZ and the MnBi part accounts for the intrinsic magnetism. In particular, the ground state for MBT$_n$ features an intrinsic A-type antiferromagnetic (AFM) order, with every layer being ferromagnetically ordered by itself while two adjacent MnBi layers carry opposite magnetizations. The AFM ground state, together with the inverted bulk band structure, makes MBT$_n$ a promising candidate for the long-sought AFM topological insulator (TI) phase  \cite{li2019intrinsic}, which is protected by an AFM TRS symmetry $\tilde{\Theta}$ that combines both the regular spinful TRS $\Theta$ and a half-unit-cell translation along [001] direction. Strong evidence of the predicted Dirac surface state has been experimentally observed via the angle-resolved photoemission spectroscopy (ARPES) technique~\cite{otrokov2019prediction}.

Applying an in-plane magnetic field along a general direction will drive the magnetic moments of MBT$_n$ to cant along the field direction. Such a canted AFM order generally breaks all internal and lattice symmetries of MBT$_n$ (including $\tilde{\Theta}$) except for the spatial inversion ${\cal I}$. The lacking of AFM TRS in the canted AFM phase trivializes the previous band topology and leads to gapped surface everywhere. Nonetheless, Zhang {\it et al.} showed that the canted AFM phase is not completely trivial, but features an inversion-protected higher-order topology with chiral ``hinge" states instead. A detailed discussion of the higher-order topology for MBT$_n$ is beyond the scope of this paper and can be found in Ref. \cite{zhang2020mobius}.

A key observation made by Zhang {\it et al.} is that when the applied field is along [100] direction, the canted AFM phase respects the same glide mirror symmetry $\hat{g}_x=\{\hat{m}_x|\hat{\tau}\}$ as the one we defined earlier in this section. The glide mirror symmetry here is clearly illustrated in Fig.~ \ref{Fig4} (c),
where $\hat{m}_x$ flips the $z$-component of the ``red spin" and changes it into a ``green spin" and the half-lattice translation sends the new ``green spin" to where it belongs to in the upper sublayer. Consequently, a gapless M\"obius surface state emerges on (010) surfaces as a result of both topological band inversion and the glide mirror symmetry in the field-induced canted AFM phase of MBT$_n$, which is numerically confirmed by an effective model calculation shown in Fig.~\ref{Fig4} (d).

Zhang {\it et al.} further realized that the front and back (010) M\"obius surfaces are not isolated, but are further connected by a pair of hinge-localized chirally propagating electron modes to form a closed loop of gapless electron channels. The existence of 1D hinge modes is confirmed through an open geometry calculation in Fig.~\ref{Fig4} (e). As discussed in Ref. \cite{zhang2020mobius}, the coexistence of 2D surface M\"obius fermions and 1D hinge chiral fermions are bulk topologically enforced and their presence must be simultaneous. Such exotic topological boundary shown in Fig.~ \ref{Fig4} (f) establishes MBT$_n$ as a novel topological state termed {\it higher-order M\"obius insulator} (HOMI). Notably, the magnetic-field generation of HOMI physics is general and not limited to MBT$_n$. For example, Ref.~\cite{zhang2020mobius} also proposed HOMI phase to occur in other AFM axion insulator candidates, such as EuIn$_2$As$_2$, EuSn$_2$As$_2$, and EuSn$_2$P$_2$. Nonetheless, MBT$_n$ remains to be the most promising platform for the HOMI realization, especially given the recent rapid experimental progress in this family of materials.

\subsection{Type-B SEBC: Hourglass Fermions}\label{Sec:hourglass}

\begin{figure*}[t]
	\includegraphics[width=0.8\linewidth]{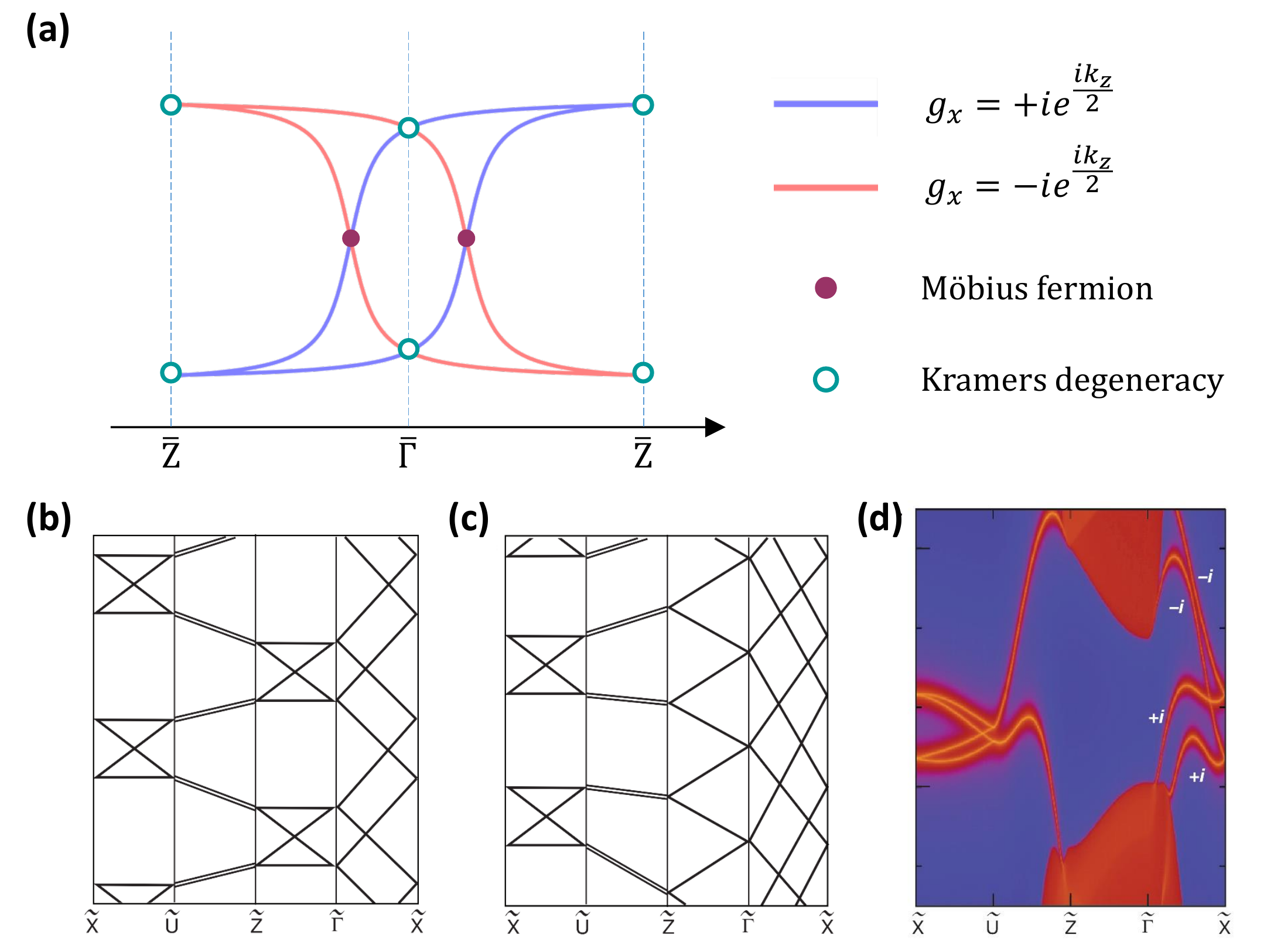}
	\caption{(a) Schematics for hourglass fermions. The color represents the glide eigen-values of the bands. The red point labels the type-B SEBC, which
		shares the same origin as that for M\"obius fermions. The green empty circles represent the Kramers degeneracies. (b) and (c) represent two
		different types of band connections of hourglass fermions. (d) The surface states spectrum for for (010) surface of KHgSb.
		(b) - (d) are from Ref.~\cite{wang2016hourglass} }
	\label{Fig5}
\end{figure*}

{\it An hourglass fermion} is a time-reversal-symmetric generalization of a M\"obius fermion, which was first proposed by Wang {\it et al.} \cite{wang2016hourglass} as the topological surface state for KHgX class of materials (X$=$As, Sb, Bi). The same hourglass-shaped surface state was later proposed to exist in certain Kondo insulators such as CeNiSn and Ce$_3$Bi$_4$Pt$_3$ as well \cite{chang2017mobius}. A schematic plot of the surface hourglass fermion dispersion is shown in Fig.~ \ref{Fig5} (a). Intuitively, an hourglass fermion is essentially a collection of two M\"obius fermions that are time-reversal conjugate to each other, which clearly manifest as a type-B SEBC. Therefore, a minimal set of glide-preserving band manifold necessarily consists of four energy bands, with two of them labeled by $g_x=+ie^{ik_z/2}$ and the other two carrying $g_x=-ie^{ik_z/2}$. Unlike M\"obius fermions, TRS also enforces the band set to form two Kramers' degeneracies at each TRS-invariant momentum of the surface BZ (i.e. $\bar{\Gamma}$ and $\bar{Z}$), as shown in Fig.~ \ref{Fig5} (a). The unique band connectivity along GIL shapes just like a pair of mirrored hourglasses, and this is how the ``hourglass fermion" got its name.

For 3D class AII systems with both spinful TRS and an additional glide mirror symmetry, K-theory predicts a $\mathbb{Z}_4$ classification~\cite{shiozaki2016topology}. A responsible topological invariant $\nu\in\mathbb{Z}_4$ can be constructed as an integral of Berry connection. The $\mathbb{Z}_4$ topological structure corresponds to a simple boundary interpretation: (i) $\nu=0$ signals a trivial phase with no surface state; (ii) $\nu=1$ signals a strong TI phase with a single Dirac surface state; (iii) $\nu=2$ signals the topological crystalline insulator phase with one surface hourglass fermion on glide-invariant surfaces (shown in Fig.~ \ref{Fig5} (b)); (iv) $\nu=3$ corresponds to a superposition of $\nu=1$ and $\nu=2$, with the coexistence of both a Dirac surface state and a surface hourglass fermion
(shown in Fig.~ \ref{Fig5} (c)). In particular, first-principles calculation by Wang {\it et al.} \cite{wang2016hourglass} predicts the KHgX family to feature
$\nu=2$ and thus a single hourglass fermion on each of its glide-invariant surface. An ab-initio-based calculation for (010) surface of KHgSb is shown in
Fig.~ \ref{Fig5} (d) to clearly demonstrate the existence of hourglass fermion state. Notably, experimental evidence for surface hourglass fermions in KHgSb was
soon observed by Ma {\it et al.} \cite{ma2017experimental} with the help of ARPES technique.

\section{Conclusion}

As a crucial ingredient of lattice space groups, nonsymmorphic crystalline symmetries not only enforce and protects exotic fermionic quasiparticles, but are also responsible for greatly enriching the topological structures of energy bands in quantum materials. We have discussed how Dirac, M\"obius, and hourglass fermions can naturally emerge in various solid-state topological crystals, as required by the underlying nonsymmorphic lattice group. Besides our target systems, other material platforms with similar physics have been extensively studied in the recent literature. For example, hourglass semimetals, a nodal band system with {\it bulk} hourglass fermionic quasiparticles, have been proposed to exist in certain 2D lattices~\cite{young2015dirac2d}. 3D nonsymmorphic lattice can support hourglass Dirac/Weyl nodal loops~\cite{bzduvsek2016nodal,wang2017hourglassPRB,wang2017hourglass} as a higher-dimensional generalization of hourglass point-nodal fermions. Remarkably, nonsymmorphic symmetry can also protect a special 4-fold-degenerate Dirac surface state (i.e. Wallpaper fermions) for topological crystalline insulators with certain lattice space groups~\cite{wieder2018wallpaper}. The above research progress unveils the surprisingly rich topological structures of electronic bands and continue to revolutionize our understanding of the century-old band theory \cite{bradlyn2017topological}.

Looking forward, the marriage between nonsymmorphic crystal symmetry and topological matter has gone beyond the context of free-fermion band systems in equilibrium. For example, straightforward generalizations to superconducting states have been achieved by multiple works, where nonsymmorphic symmetry is found to protect Dirac~\cite{wang2016topological}, M\"obius~\cite{yanase2017mobius}, and hourglass-like~\cite{daido2019z} Majorana boundary modes. New nonsymmorphic symmetries on a spacetime lattice have also been proposed for periodic driven Floquet dynamical systems, which generally consist of reflection or rotation operations on the spatial coordinates followed by a fractional translation along the time direction. These operations are termed ``time-glide" and ``time-screw" symmetries and are responsible for protecting novel dynamical topological matters that have no static counterpart~\cite{morimoto2017floquet,peng2019floquet}. Nonsymmorphic symmetry is also vital for interacting electron physics, and it was shown that nonsymmorphic symmetry can forbid band insulator at certain integer fillings for interacting electrons in solids, and if the system shows
insulating behavior in this case, it must exhibit fractionalized excitations and topological order \cite{parameswaran2013topological}. 
Beyond condensed matter systems, nonsymmorphic-symmetry-protected topological phases have also been explored in photonic systems
\cite{xia2019observation} and cold atom systems \cite{lang2017nodal,yan2019emergent}. It will not be surprising that our deepening
understanding of both symmetry and topology in condensed matter physics may lead to new applications of nonsymmorphic symmetry 
in the topological matter in the future. \\

\begin{acknowledgments}
	The authors are honored to contribute to the Festschrift in Honor of the C N Yang Centenary.
\end{acknowledgments}

\bibliographystyle{apsrev4-2}
\bibliography{ref}

\end{document}